\documentclass[preprint,1p,times]{elsarticle}
\journal{}

\usepackage{threeparttable}
\usepackage{multirow}
\usepackage{bbm}
\usepackage{algpseudocode}

\usepackage{xspace}
\usepackage{ulem}
\usepackage{cancel}

\usepackage{algorithm2e}
\usepackage{xcolor}
\usepackage[utf8]{inputenc} 
\usepackage[T1]{fontenc}    
\usepackage{hyperref}       
\usepackage{url}            
\usepackage{booktabs}       
\usepackage{amsfonts}       
\usepackage{nicefrac}       
\usepackage{microtype}      
\usepackage{bm}     
\usepackage{upgreek}
\usepackage{mathtools}
\usepackage{tensor}
\usepackage{color}
\usepackage{wasysym}
\usepackage{kpfonts}
\usepackage{todonotes}
\usepackage{amssymb}
\usepackage{amsthm}
\usepackage[bbgreekl]{mathbbol}
\usepackage{amsmath}
\usepackage{cleveref}
\usepackage{epigraph}
\usepackage{ stmaryrd }
\usepackage{scalerel}
\usepackage{stackengine}
\DeclareSymbolFontAlphabet{\mathbbm}{bbold}
\DeclareSymbolFontAlphabet{\mathbb}{AMSb}

\theoremstyle{definition}

\theoremstyle{definition}

\theoremstyle{definition}

\theoremstyle{definition}

\include{op-aux}
\newcommand{\shorten}[2]{%
  \mathop{
    \sbox0{$\displaystyle#1$}
    \raisebox{-\height+\ht0}[\ht0][\dp0]{\scalebox{#2}[1]{\copy0}}
  }\displaylimits
}

\newcommand{\con}{\shorten{\hspace{-1pt}\frown\hspace{1pt}}{0.6}}

\newcommand{\ord}{\omega}
\newcommand{\minord}{\Omega}
\newcommand{\ph}[1]{\overset{\times}{#1}}
\newcommand{\inv}[1]{#1^{-1}}
\newcommand{\idx}[1]{\rho_{\texttt{+}#1}}
\newcommand{\g}[1]{\bm{\mathrm{UT}}^{4}_{#1}}
\newcommand{\G}[1]{\bm{\mathrm{UT}}^{5}_{#1}}

\definecolor{dark-green}{rgb}{0,0.5,0}
\newcommand{\green}[1]{\textcolor{dark-green}{#1}\xspace}
\newcommand{\come}[1]{\green{\scriptsize\phantom{i}$\triangleleft$\phantom{i}\textit{#1}\xspace}}

\newcommand{\set}[1]{\{ #1 \}}
\newcommand{\tup}[1]{\langle #1 \rangle}

\theoremstyle{definition}

\usepackage{longfbox,varwidth}
\newcommand{\hs}[1]{
    \lfbox[border-color=lightgray, rounded,tight,padding={1pt,1pt}]{
        \footnotesize
        \textcolor{darkgray}{\texttt{#1\vphantom{I}}}
    }\xspace
}

\newcommand{\half}{{-}{-}{-}{-}{-}{-}{-}{-}{-}{-}{-}{-}{-}{-}{-}{-}{-}{-}{-}{-}}
\newcommand{\halfp}{{-}{-}{-}{-}{-}{-}{-}{-}{-}{-}{-}{-}{-}{-}{-}{-}{-}{-}{-}{0}}

\newcommand{\q}[1]{    
    \lfbox[border-color=lightgray,tight,padding={0.6pt,-2.5pt}]{        
        \textcolor{darkgray}{\texttt{#1\vphantom{I}}}
    }\xspace
}

\newcommand{\del}[1]{
    \delta_\text{\hspace{-2pt}\q{\footnotesize#1}}
}

\newcommand{\grp}[1]{\bm{\mathrm{#1}}}
\newcommand{\objs}{\mathfrak{O}}

\newcommand{\tex}[1]{
    \text{\hspace{-2pt}\q{\footnotesize#1}}
}

\newcommand{\zfill}{\_\_\_\_\_\_\_\_\_\_\_\_\_\_\_\_\_\_\_\_\_\_\_\_\_\_\_\_\_}
  
\usepackage{textcomp}
\usepackage{trimclip}
\makeatletter
\DeclareRobustCommand{\shortto}{%
  \mathrel{\mathpalette\short@to\relax}%
}
\newcommand{\short@to}[2]{%
  \mkern2mu
  \clipbox{{.2\width} 0 0 0}{$\m@th#1\vphantom{+}{\shortrightarrow}$}%
  }
\makeatother

\begin{document}

\begin{frontmatter}
\title{Predictable universally unique identification of sequential events on complex objects}

\author{Davi Pereira-Santos}
\ead{dpsbac@gmail.com}
\cortext[cor1]{Corresponding author}

\author{Gabriel Dalforno Silvestre} 
\ead{gdalforno7@usp.br}

\author{Andr\'e C. P. L. F. Carvalho} 
\ead{andre@icmc.usp.br}

\address{
    Instituto de Ci\^encias Matem\'aticas e de Computa\c{c}\~ao, Universidade de S\~ao Paulo,
    \\ Trabalhador S\~ao-carlense Av. 400,
    S\~ao Carlos,
    13560-970,
    S\~ao Paulo,
    Brazil
}

\begin{abstract}
Universal identifiers and hashing have been widely adopted in computer science from distributed financial transactions to data science. %
This is a consequence of their capability to avoid many shortcomings of relative identifiers, such as limited scope and the need for central management. %
However, the current identifiers in use are isolated entities which cannot provide much information about the relationship between objects. %
As an example, if one has both the identifiers of an object and its transformed version, no information about how they are related can be obtained without resorting to an authority or additionally appended information. %
Moreover, given an input object and an arbitrarily long sequence of costly steps, one cannot currently predict the identifier of the outcome without actually processing the entire sequence. %
The capability of predicting the resulting identifier and avoiding redundant calculations is highly desirable in an efficient unmanaged system. %
In this paper, we propose a new kind of unique identifier which is calculated from the list of events that can produce an object, instead of directly identifying it by content. %
This provides an identification scheme regardless of the object's current existence,
thus allowing to inexpensively check for its content in a database and retrieve it when it has already been calculated before. %
We propose these identifiers in the context of abstract algebra, where objects are represented by elements that can be operated according to useful properties, such as
associativity, order sensitivity when desired, and reversibility, while simplicity of implementation and strong guarantees are given by well-known group theory results.
\end{abstract}




\end{frontmatter}

\epigraph{If you don't know where you want to go, then it doesn't matter which path you take.}{Lewis Carroll, Alice in Wonderland}

\section{Introduction}\label{sec:intro}
From the scientific binomial nomenclature \cite{cracraft1983species} to national databases of personal identification numbers, rigorous naming of objects or classes thereof is an essential step when one intends to model some part of reality. %
For the sake of clarity, in this text\footnote{Here, we focus on universal identifiers, i.e., those which are valid as identity within any scope.}, we will refer to such attributed names and respective entities as \textit{identifiers} and \textit{objects}.
Ideally, the possession of an identifier would be equivalent to  direct access to any information of interest about its related object. %
However, there is a trade-off related to the intended degree of abstraction/simplification. %
Perfectly general mapping, i.e., one that serves for all possible purposes, would contain all information about the object embedded in the identifier. %
In this contrived case, the object would need to be represented by itself. %
Alternatively, in a completely digital context, a lossless compression of the object, when possible, could be used as such a perfect identifier. %
At the opposite extreme, hash digests or simple nonordinal identification numbers just show whether two objects are the same or not. %
This latter extreme would require a companion database which one needs to resort to if any relevant information about the object is needed. %
Some possibilities that lie between these extremes are enumerated in \Cref{sec:related} for different amounts of embedded information. %

Ideally, fetching data from a distributed unmanaged environment would only need a single identifier. %
This identifier is the key to be provided when querying databases in the network. %
However, if the desired object is necessarily a result of some calculation, it is impossible to directly know its deterministic identity beforehand without actually performing the calculation. %
Traditionally, this is solved by providing the identities of all parts involved in the output object generation. %
A simple example is to assign an identifier to the calculation process and another to the input object identifier. %
This pair of identifiers acts as a composite key to query relational databases. %
In such a setting, the database needs an internal structure that takes into account the different entities involved. %
Additional complexity is present also from the user interface perspective due to the different possible amounts of identifiers contained in a query: %
the output object identity could be provided directly; or, in its absence, the input object and its transformation process identities could be provided instead. %
Moreover, this solution is not scalable with respect to the number of steps that can sequentially transform the object, as a variable-length list of identifiers would have to be included in the query. %
A traditional scalable solution requires a more sophisticated database structure and querying interface, departing from the concept of unmanaged environments - at least in the sense of freedom and simplicity of implementation. %

In this article, we propose a uniform identification approach for data and processes. %
A typical storage for this proposal is as simple as key-value engines, e.g., NoSQL \cite{han2011survey} or relational databases with a single table. %
Both entities (data and process) are seen as identified objects, where the amount of values or steps inside each object can be arbitrary. %
In this scenario, there is no distinction between data and process from the identification perspective as any data object results from a process. %
We build upon this assumption to set up convenient relationships between identifiers in different levels during data transformation. %
For this purpose, we adopt abstract algebra to operate over identifiers as part of a \textit{group}. %
Consequently, well established mathematical group properties can operate identifiers with sound guarantees against ambiguity when needed. %
Applicability of the method is vast, including data structure libraries, data science and current mainstream technologies such as \textit{content addressable storages} \cite{ratnasamy2001scalable,van2016data}. %
Besides solving the previously mentioned issues, the proposed approach inherently offers useful additional functionalities that bring identity handling to a new level of flexibility. %
The concepts presented here were implemented in a software package\footnote{\url{https://pypi.org/project/garoupa}} \cite{davi_pereira_dos_santos_2021_5501845}. %

Prominent current identification schemes and related work about abstract algebra are presented in \Cref{sec:related}. %
A brief overview, the intended scenario of application (which is also part of our contribution), its requirements, terminology, and notation are given in \Cref{sec:overview}. %
The proposed method is defined in \Cref{sec:method}. %
Finally, a brief comparison between the properties of alternative abstract algebra groups is presented in \Cref{sec:comparison}, limitations of our approach in \Cref{sec:lim} and future work in \Cref{sec:future}. %

\section{Related Work}\label{sec:related}
Most of the existing work on data identification is related to the identifier presentation, where semantics is often limited to the represented object, i.e., there is little or no embedded information linking objects within the identifier. %
Another type of related work includes groups from abstract algebra suitable to be used as sets of identifiers that can be combined. %
Both types are briefly reviewed in the next two sections. %

\subsection{Representation of identity}
Different identifier types contain varying amounts of embedded information. %
They can be one of the extremes mentioned in \Cref{sec:intro} or have an intermediate kind. %
For instance, the \textit{binomial nomenclature} provides information about the taxonomic hierarchy of organisms with descriptive Latin words \cite{cracraft1983species}. %
These words can bring additional information like object appearance (species phenotype), features or some tribute related to the object history (species discovery), among others. %
Similarly, standard universally unique identifiers (UUID) embed different kinds and degrees of information, depending on the specification version \cite{uuid}: %
time and/or local hardware attributes; %
random or pseudo-random numbers; %
and, cryptographic hash of the content of objects. %
We are specifically interested in hash-based identifiers (such as UUID versions 3 and 5), because, for a given object, they are deterministic, i.e., independent of time and place of generation. %

Another type of identification is provided by \textit{access tokens}, which are usually longer and contain information about user accounts and access permissions to a given remote service \cite{sandhu1998role}.
We will focus on less volatile identifiers. %
ORCID and DOI \cite{haak2012orcid,paskin2010digital} are examples of managed permanent identification systems for researchers and publications. %
One application of our proposal is to have some degree of interoperability with these established managed systems and also with hash-based systems, e.g., git version control \cite{chacon2014pro}. %

\subsection{Abstract algebra}\label{sec:groups}
The chronological nature of data processing can be exactly represented by an expression of juxtaposed identifiers. %
This sort of syntax appear in several other contexts such as stack-based processors, reverse Polish notation and concatenative programming \cite{herzberg2009concatenative}. %
A major advantage of using expressions is that they can be reduced, combined or manipulated according to predefined laws. %
Therefore, abstract algebra \textit{groups} can fit exactly the intended task of flexible identification of objects, provided the group allows arbitrarily long sequences of operations - including repetitions. %
While we could not find any studies on abstract algebra directly focused on our intended setting, adoption of finite groups for hashing is not new. %
\textit{Incremental hash} is the closest task found in the literature for operating identifiers. %
This type of hash is more general than needed here, as it operates text segments of size not previously defined. %
A very straightforward implementation for the incremental hashes is the adoption of the additive group of integers modulo $p$, where $p$ is prime \cite{bellare1994incremental}. %
However, it is a commutative solution, which is inadequate to unambiguously represent sequences of identifiers. %
For instance, consecutive steps could be represented as a single element resulting from the group operation applied over their identifiers, but due to the commutative nature of addition, elements are order-insensitive. %

On the other hand, finite non-Abelian groups are suitable for our task as long as the commutativity degree is low enough or manageable within subgroups \cite{castelaz2010commutativity}. %
Some examples of non-Abelian finite groups are: symmetric, dihedral, quartenion, unitriangular matrix, and wreath product of cyclic groups, among others. %

\section{Overview}\label{sec:overview}
\begin{figure}
\centering
 \includegraphics[width=10cm]{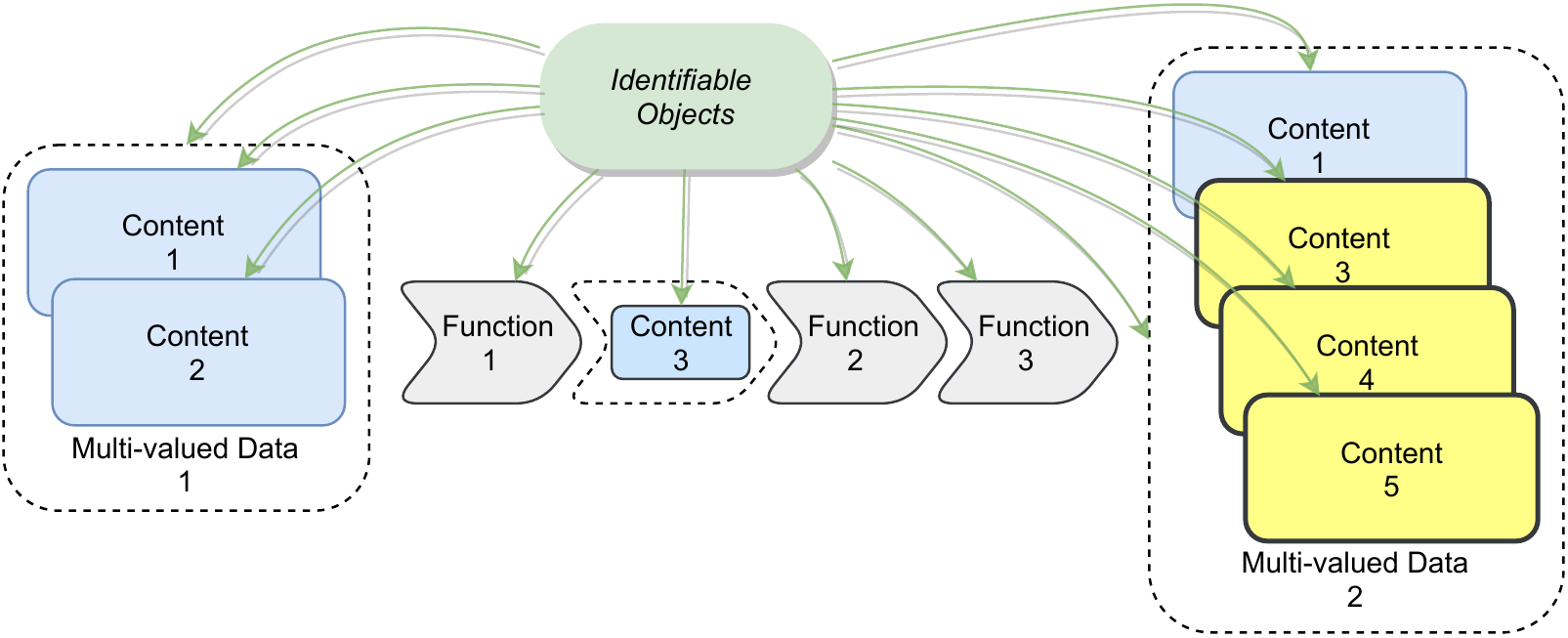}
 \caption{Multi-valued data processing is our motivating example of application. In this scenario, the relationships between identities define both the process and the content outcome.}
\label{fig:caixas}
\end{figure}

The broader context of this work is any application where the combination of an arbitrary number of identifiers is meaningful. %
The combination can refer to a variety of meanings - from a simple unordered collection of objects to a sequence of events in time. %
As a motivating example application, we adopted the common and difficult task of identity attribution to multiple-step multi-valued data processes and outcomes throughout the text, as shown in \Cref{fig:caixas}. %
A collection of values is provided as input to a sequence of functions to be reduced, extended or internally changed. %
We consider the three-way identification described as follows as a requirement of our intended target setting. %

A multi-valued data object and the process that can generate it are two equivalent sources of identity. %
This is possible because the concept of \textit{process} always includes the data generation from the beginning, i.e., even the inclusion of the first original value is part of the process. %
Therefore, the same identifier is obtained from either \textit{content} or \textit{process} information. %
The third equivalent source of identification is given by the \textit{history} of irreducible operations that generated the data object (\Cref{sec:hist}). %
Process and history steps often coincide. %
Briefly, an identifier can be calculated from algebraic expressions (\Cref{sec:notation}) generated under three different perspectives:
\begin{itemize}
    \item process - general sequence of identifiers representing the steps intended by the user; %
    \item data - sequence of identifiers representing values; and, %
    \item history - detailed sequence of identifiers representing the actually performed (atomic) steps. %
\end{itemize}

\subsection{Terminology}\label{sec:term}
Information is often organized into multiple parts in many contexts. %
We adopt a collection of \textit{values} to this end, e.g., tuples. %
Each processing step is a \textit{function} that transforms a collection of values. %
For the sake of simplicity, we often refer to the collection as a tuple, but the collection can have another type of structure such as a key-value map or a set. %
There is no distinction between an original (i.e., non-transformed) value and the function that inserts that value at the identification level. %
This is a consequence of the three-way identification mentioned at the beginning of the \Cref{sec:overview}, i.e., a value and its insertion function have the same identifier. %
However, in this text, we make the distinction between value and function identifiers to be able to provide more meaningful examples. %
Both entities are identifiable \textit{objects} represented in our proposed abstract algebra group by \textit{value-elements} and \textit{function-elements}. %
The adopted notation consider all objects as functions, indistinguishably (\Cref{sec:notation}). %

An element has either \textit{original} or \textit{composite} nature. %
Singletons (i.e., $1$-tuples) and original functions are similar in the sense that they are atomic objects, i.e., their respective elements are not a result of group operations, because they do not represent a combination of objects. %
For instance, a function written by a programmer or a file with temperature measurements are original objects. %
Their corresponding group elements would be generated directly from a hashing algorithm, or arbitrarily provided by the programmer. %
On the other hand, a composite element would be generated when the file is transformed by a function that converts Celsius degrees to Fahrenheit. %
In this case, the resulting value-element is generated by the group operation over the identifiers of the function and the file. %
This example is expressed as an algebraic expression in the \Cref{sec:notation}. %
Finally, a \textit{removal} identifier represents the removal of a value by its position in a tuple, or by its name/key on a map. %

\subsection{Notation}\label{sec:notation}
Given the finite set $\grp{U}$ of identifiers and set $\objs$ of objects, the function $\varphi\colon\grp{U}\to\objs$ maps identifiers to objects.
This function can be considered surjective for practical purposes and is hypothetically defined here just for convenience and is actually an impossible\footnote{ %
    Computationally intractable (or unsolvable, when collisions are taken into account). %
} function because it represents hash reversion of original elements and reverse mapping for composite elements. %
Moreover, for the sake of simplicity, we call hash the function $\gamma\colon\objs\to\grp{U}$ defined as $\gamma=\inv{\varphi}$. %
The disjoint union %
$\grp{U} = \mathfrak{V}\sqcup \mathfrak{F}\sqcup \set{\phi}$ represents different categories of identifiable objects, respectively bound to groups of adjacent letters from the Latin alphabet for convenience, except for $\phi$ as a special reserved symbol: %
\begin{itemize}
 \item General elements %
    \subitem $a,b,c,d \in \grp{U}$;
 \item Singletons, larger tuples and removed values %
    ($\mathfrak{V}=\mathfrak{V}^1\sqcup\mathfrak{V}^{2,3\cdots}\sqcup\mathfrak{V}^\times$) %
    \subitem $x,y,z,w \in \mathfrak{V}^1$; %
    \subitem $u,v \in \mathfrak{V}^{2,3\cdots}$; %
    \subitem $\ph{x},\ph{y},\ph{z},\ph{w} \in \mathfrak{V}^\times \implies x,y,z,w \in \mathfrak{V}^1$; %
 \item Original and composite functions ($\mathfrak{F}=\mathfrak{F}^1\sqcup\mathfrak{F}^{2,3\cdots}$) %
    \subitem $f,g,h \in \mathfrak{F}^1$; and, %
    \subitem $s \in \mathfrak{F}^{2,3\cdots}$. %
 \item Empty tuple/Identity function/Zero element
    \subitem $\phi \in \grp{U}$;
\end{itemize}

Resuming the example from \Cref{sec:term}, the conversion of temperature data in a file identified by $x$ from Celsius to Fahrenheit through the function identified by $f$ would be represented by $xf$. %
The expression $xf$ represents the application of the function $\varphi(f)$ to the value $\varphi(x)$.
If other functions were to be applied, such as $\varphi(g)$ followed by the insertion of an additional value $\varphi(y)$, the resulting identifier would be represented by the sequence of group operations $xfgy$. %
Similarly, a multi-valued data $\varphi(u) = \tup{\varphi(x),\varphi(y)}$ is also represented by a product of identifiers: %
$u = xy$. 
Tuple identifiers are operable by function-elements, similar to any other value-element: %
$uf = xyf$. %
When the function generates a new value $\varphi(z)$, it extends the input tuple. %
We derive a formula for adding new identifiers ($z$ in the example) which is consistent with the identifier of the resulting tuple (\Cref{sec:multi}). %
Finally, some additional definitions are provided in the next sections. %

\subsubsection{Group and order}\label{sec:gorder}
Given a set $\grp{G}$ of identifiers and an operator $\cdot$, we refer to the group $\tup{\grp{G},\cdot}$ just as $\grp{G}$. %
The minimum non-trivial element order (\Cref{sec:eorder}) of $\grp{G}$ is represented by $\minord(\grp{G})$, and the order of an element $a\in\grp{G}$ is given by $\ord(a)$. %
The symbols $\grp{Z}$ and $\grp{H}$ are, respectively, the center and the maximal Abelian subgroup of $\grp{G}$. %
In this context, two properties on commutativity are convenient to represent unordered values: %
$am=ma \forall a\in\grp{G}, m\in\grp{Z}$; and, %
$xy=yx \forall x,y\in\grp{H}$.

\subsubsection{Identity element}
The identity element $\phi$ represents both the empty tuple and the reflexive\footnote{
    Here, the term \textit{reflexive function} is preferred over \textit{identity function} to avoid confusion with the     \textit{identity element} of a group.
} function, since neither of them change the tuple. %
In other words, when $\varphi(\phi)$ is considered as a null value, $xy\phi$ represents the inclusion of nothing into a tuple, whereas when $\varphi(\phi)$ is considered as the reflexive function %
$r(o)=o, o\in\mathfrak{O}$, the expression $xy\phi$ represents the application of the function $r$. %
In both cases, the result is the same at the object level: %
$\tup{\varphi(x),\varphi(y)}\con\tup{\phantom{|}} = r(\tup{\varphi(x),\varphi(y)}) = \tup{\varphi(x),\varphi(y)}$. %

\subsubsection{Tuple-expression equivalence}
Given a finite sequence of $k$ elements $a_1,a_2,a_3,\cdots,a_k \in \grp{U}$, we define the \textit{tuple-expression equivalence} as $\tup{a_1,a_2,a_3,\cdots,a_k} \equiv a_1a_2a_3\cdots a_k$. %
This equivalence comes from the factual equality $\varphi(\tup{x,y}) = \tup{\varphi(x),\varphi(y)}$ implicit in the relationship between value objects - consequence of the three-way identification requirement (beginning of \Cref{sec:overview}). %
Additionally, the equivalence is convenient when dealing with the Cartesian product between two or more sets of identifiers (\Cref{sec:par}), allowing a direct map between tuples and expressions. %

\subsubsection{Removal identifier}\label{sec:removal}
A removal identifier $\delta_i$ (or, $\delta_\eta$ for maps) is expected to be provided as a means to represent the removal of a value by its position $i$ in the data tuple (or, by its name $\eta$ in a map). %
For instance, when $\delta_2$ (or $\del{y}$) is applied to an expression $xyz$, a placeholder $\varphi(\ph{y})$ is created in the resulting tuple: %
$\tup{\varphi(x),\varphi(\ph{y}),\varphi(z)}$. %
Therefore, $xyz\delta_2 = xyz\del{y} = x\ph{y}z$, where $\q{y}$ is the name associated to the value $\varphi(y)$ in a map, and $\ph{y}$ represents its absence. %
As a consequence, $\varphi(\ph{y})$ is an object with no content,
i.e., $\varphi(\ph{y}) = \varphi(\phi)$. %
Such a placeholder is needed for consistency between the expression and its represented output object as explained in \Cref{sec:deli}. %

\subsection{Compatibility}\label{sec:compatibility}
Compatibility is an important requirement for wide adoption of any identification scheme as it allows for direct data exchange between new and legacy systems. %
A current \textit{de facto} standard is the 40-digit hexadecimal digest. %
Often called hexdigests, they are base-16 identifiers composed only by numbers and lower case letters up to the letter \textit{f}. %
We adopt a specific typeface for them, e.g.: %
\hs{ea035db8b34a60ed5cbeaf568672a8f68aa1a39b}. %
Hashing algorithms, for instance SHA-1 \cite{wang2005cryptanalysis, alkandari2013cryptographic}, have been used to generate hexdigests, and are capable of uniquely identifying an object for most practical purposes \cite{patarin2005benes}. %

Textual representation is convenient for digests because it is limited to characters that do not lead to encoding incompatibility across different software and hardware platforms. %
These characters are explicitly allowed by official specification, and considered safe for use in uniform resource identifiers (URI) \cite{berners1998rfc2396}. %
URIs are a good example of critical application regarding compatibility as they are at the core of Internet communication. %
Apart from letters and numbers, some extra characters are considered safe by the specification:\hs{-},\hs{.}, \hs{\_} and \hs{\~{}}. %
Unfortunately, the character\hs{\~{}} becomes \textit{percent-encoded} in some contexts \cite{berners1998uniform}. %
This amounts to a total of 65 safe characters, allowing for a base-64 alphabet, which is convenient to represent identifiers from a large set. %
Therefore, we define the representational limit of base-64 as a requirement for our intended setting. %
It represents more objects than the 40-digit hexadecimal digest format without incurring in extra digits. %
We take advantage of such a property combined with the fact that the base-16 alphabet is a subset of the base-64 alphabet, as shown in \Cref{sec:id}. %

\subsection{Robustness}\label{sec:req}
Each abstract algebra group has particularities that can affect the feasibility of the identification system. %
Therefore, we define a common set of requirements for the candidate group in the next sections. %
Except for the computational cost, other requirements are expressed in terms of iterations until failure. %
Probability of failure is given when applicable. %
For instance, on average, an unmalicious collision occurs in 128-bit hashing algorithms such as MD5 \cite{rivest1992md5} after generating about $2.2\cdot 10^{19}$ hashes\footnote{ %
According to the approximate formula for the birthday attack for 128 bits: $P_n \approx 1-e^{(-n^2+n) / 2^{129}}$. %
}, considering a perfect \textit{balance} \cite{bellare2004hash}.
Conversely, the collision probability of a random pair is $2^{-128} \approx 2.9\cdot 10^{-39}$. %

Finally, cryptographic properties are not considered as requirements here. %
While they may be found to hold to some extent depending on the application, we anticipate that the identity of an applied function, depending on the group and elements choice, is not guaranteed to be completely hidden in the resulting element. %
For instance, an operation between elements $a\in\grp{H}$ and $b\notin\grp{H}$ results in an element $ab\notin\grp{H}$ whose identifier is partly identical to the identifier of $a$ due to the fact that $\grp{H}$ is a group by itself. %
As a consequence, operations involving its elements can only affect the part of the identifier that refers to $\grp{H}$, leaving the remaining digits unchanged. %

\subsubsection{Commutability}\label{sec:comm}
In order to represent a sequence of events unambiguously, the operation over identifiers in the candidate group $\grp{G}$ must not commute. %
Although the odds of sampling the same pair of elements twice and using each of them in a distinct ordering is sufficiently remote when they come from a uniform distribution, it is a matter of concern when the set of effectively used identifiers is very small relatively to $|\grp{G}|$. %
This is more noticeable for function-elements as functions are expected to be reused, often as part of a long composition. %
Such a reduced set implies the existence of recurring elements, which increases the odds of a sub-expression $ab\in\grp{G}$ and its reverse $ba$ being used in the same application. %
In this case, the ambiguity arises when $\varphi(ab)$ and $\varphi(ba)$ are different objects. %
Hence, absence of ambiguity among pairs is important to ensure that any expression, regardless of length, is also unambiguous, e.g., $abc\neq acb\neq \cdots\neq cba$. %
Similarly, a sequence with more than two identifiers is ambiguous if it contains at least one explicit or implicit operation that commutes. %
For instance, the expression $abc$ has two implicit elements $c,d\in\grp{G}$: %
$c=ab$ and $d=bc$. %
Despite being implicit, they should also not commute with any adjacent element in the expression. %

In the general case, the number of implicit and explicit operations in a sequence is $\binom{l+1}{3}$. %
Let $P_c(\grp{G})$ be the commuting probability of a random pair sampled from $\grp{G}$. %
The probability $P_m$ of creating an ambiguous random expression with size $l$ is given by \Cref{eq:ambexp}. %
\begin{equation}\label{eq:ambexp}
\begin{aligned}[b]
    P_m(\grp{G},l) &= \min\left\{
        1,~
        P_c(\grp{G})\cdot\binom{l+1}{3}
        \right\}
\end{aligned}
\end{equation}
Therefore, on average, an ambiguous expression is expected after $n$ samples given by \Cref{eq:nambexp}, where the complementary probability is $\overline P_m(\grp{G},l) = 1 - P_m(\grp{G},l)$. %
\begin{equation}\label{eq:nambexp}
\begin{aligned}[b]
     \\
    \overline P(\hs{none of n events}) &=1-\overline P_m(\grp{G},l)^n=0.5\implies
    n = \underset{\overline P_m(\grp{G},l)}{\log}~0.5
\end{aligned}
\end{equation}

\subsubsection{Element Order}\label{sec:eorder}
The \textit{order} of a group element defines the limit on the number of times it can be operated with itself before the result becomes the identity element. %
For a practical extreme example of repetition, %
let $\varphi(g), g\in\mathfrak{F}^1$ be a function with the only purpose of incrementing a value object to count how many times a function $\varphi(f), f\in\mathfrak{F}^1$ is applied to a tuple $\varphi(u), u\in\mathfrak{V}^{2,3,\cdots}$. %
This leads to the expression $ufg$ at the first iteration - as expected. %
After $t$ iterations the expression becomes $u[fg]^t$. %
The inequality $t<\ord(fg)$ must be respected, otherwise the expression becomes ambiguous: $u[fg]^t = u[fg]^{t\bmod \ord(fg)}$. %
Ideally, the minimum non-trivial element order $\minord(\grp{G})$ would be large enough to make it impossible to overflow the limit imposed by the order $\ord$ of an element in a real application. %
Depending on the application, when there are none or few repetitions, very small $\minord(\grp{G})$ values are still suitable. %

\subsubsection{Compatibility Gap}\label{sec:compag}
Many groups do not fit into the representational limits imposed by the current binary-oriented hardware. %
We define $\xi_\beta(\grp{G}) = 1 - |\grp{G}|\cdot 2^{-\beta}$ as the compatibility gap of a group $\grp{G}$ in relation to the expected size $2^\beta$, which is the representational limit of $\beta$ bits. %
The gap can be avoided through representational constraints at the digest level (\textit{compatibility trick} in \Cref{sec:id}); at the group level via direct product to reduce it; or, probabilistically, if $\xi_\beta(\grp{G})$ is negligible. %

\subsubsection{Computational Cost}\label{sec:compu}
Multi-valued data is usually handled in memory as structures such as lookup tables or lists. %
The extra processing overhead added by the identification handling should be negligible when compared with the actual core data processing in a given task. %
We would not expect the operation over identifiers to have a significant difference in processing time when compared, e.g., to a conventional hash applied to the structured content. %
This is a conservative reference, but reasonable, as both current and proposed methods aim to provide identification - despite the broader context and flexibility of the setting we address. %
As a conservative reference, hashing small texts through MD5 takes approximately $1.6\mu s$ in current hardware\footnote{
    Time measured for hashing 1kiB of data using Python 3.8.5 and processor Intel i7-8565U at 1.80GHz.
}. %
A more realistic reference would consider the computational cost of constantly hashing potentially large contents after each data modification. %
In such a case, the cost of an algebra-based solution is comparatively negligible. %

\section{Proposed Method}\label{sec:method}
We propose a partly hash-based UUID that includes the definition of an operation to combine an arbitrary number of (potentially ordered) identifiers. %
The use of a hashing algorithm from the literature for original values, and the adoption of a group from abstract algebra fulfill this goal and enables appealing incidental possibilities, such as reversibility and associativity. %
The format of the identifier provided as input to the operation is backwards-compatible with current UUID specification as long as the identifier is represented in the standard hexadecimal format. %
Similarly, the output from the operation can be directly fed into systems that handle or store standard UUIDs as text. %

Unfortunately, the groups whose size fits all elements exactly into a binary representation behave poorly regarding commutativity and/or repetition of operations. %
Therefore, we propose a simple \textit{compatibility trick} to overcome the need for a perfect match between group order and binary representation. %
It can keep the digest size within the same number of digits even though the number of bits is larger. %
This increases the group order, improving its robustness to viable levels. %
We consider using 160 bits the most recommended choice, targeting compatibility with 40-digit hashes as shown in \Cref{sec:id}. %
We name it UT40.4 and propose it as a parameterized UUID version referring to: the chosen group (UT); the size of the digest in digits (40); and, the value (4) of a specific parameter of the unitriangular matrix group (details in \Cref{sec:group}). %
For the sake of  brevity, the examples in this article are focused on version UT40.4, intended to be compatible with most systems in use today. %
We provide a reference implementation containing also two additional versions with distinct digest sizes: UT32.4 and UT64.4 \cite{davi_pereira_dos_santos_2021_5501845}. %

\subsection{Identifier choice}\label{sec:id}
We propose a trick to address the need for a larger identifier space without losing compatibility. %
Its principle is to consider textual digests the only legitimate type of representation of an identifier, ignoring how it would be expressed as bytes or as a number. %
This brings independence from the actual memory size/disposition needed to store it. %
Therefore, while a usual 160-bit hexadecimal digest is represented by 40 digits from the alphabet \hs{0123456789abcdef}, the same 40 digits could represent a 240-bit base-64 digest using the alphabet \hs{0123456789abcdefghijklmnopqrstuvwxyzABCDEFGHIJKLMNOPQRSTUVWXYZ-.}. %
Besides allowing data exchange from/to hexadecimal systems, the greater number of bits the trick provides prompts for a larger and, therefore, more robust group. %
At the binary level, the chosen group has a size that fits within 30 bytes. %
Different bases for each interval of textual identifiers are defined  for the sake of convenience when importing identifiers from legacy systems (\ref{appendix}). %
In short, the proposed identification scheme supports as input the following intervals of identifiers, where $p=2^{40}\texttt{-}87$ is a prime number corresponding to the UT40.4 version. %
\begin{itemize}
  \item All base-16 identifiers (most used); %
  \item All base-62 identifiers (often used); %
  \item Additionally: %
    \subitem All 239-bit identifiers, i.e., 30 bytes with zero at the most significant bit; %
    \subitem All 39-digit base-64 identifiers, i.e., the most significant digit is \hs{0}; %
  \item In the limit: %
     \subitem All base-64 identifiers in the interval $[0; p^6-1]$;
  \item Special cases (more about order-sensitivity in \Cref{sec:abel}): %
    \subitem Operating an identifier from the interval $[0; p-1]$ is order-insensitive; %
    \subitem Operations within the interval $[0; p^4-1]$ are order-insensitive; and, %
    \subitem Reserved prefixed identifiers, described as follows. %
\end{itemize}

Elements from the larger order-insensitive subgroup only commute among themselves. %
Their number amounts to about $2^{160}$ for version UT40.4 which is conveniently representable by 40 hexadecimal digits. %
Thus, multiple identifier can be imported from legacy systems and operated without imposing any order on them. %
While the proposed group accepts many input formats, the result of any further operation will always produce 240-bit base-64 identifiers. %

The identifier $\rho=\hs{\half\halfp}$ and subsequent lexicographic elements are reserved to decompose a value-element resulting from the application of a function that returns multiple values (\Cref{sec:multi}). %
Other identifiers, prefixed by \hs{\half}, are reserved for removal identifiers (\Cref{sec:notation}). %
They are padded with repetitions of the digit \hs{{.}} until the start of a numeric index or an alphanumeric name. %

Original value-elements are generated by the first 239 bits of BLAKE3 hashing algorithm \cite{blake3} modulo $p^4$. %
Any other algorithm could be used, but we specify one here as a reference to ensure all original values will produce consistent identifiers across different systems. %
Additionally, BLAKE3 is multithreaded and among the faster cryptographic hashing algorithms currently. %
According to its authors, it is considered safe for all security purposes, despite the adoption of less rounds than its conservative predecessor \cite{aumasson2013blake2}. %
Function-elements can be generated in the same way as value-elements, i.e., through BLAKE3 hashing, except by not applying modulo $p^4$. %
The binary content can be, e.g., the abstract syntax tree of the implemented function, its bytecode, or a binary string containing a fully qualified name of the function along with the parameters used. %
The uniqueness of function-elements depends on this choice. %

Software systems that require an identification scheme are often ported to different languages and platforms. %
In this case, all versions of a function can, and should, certainly share the same identifier, provided one can ensure they perform the exact same transformation on data. %

\subsection{Group choice}\label{sec:group}
After considering the properties of all groups enumerated in \Cref{sec:groups}, we opted for a unitriangular matrix group \cite{weir1955sylow}. %
Ideally, the group order would be a power of $2^8$ or, at least, a power of $2^1$, due to the representational limitation imposed by computational bytes and bits, respectively. %
However, commutativity and element orders are compromised in such groups when compared to groups of odd order as far as we can tell from the studied groups (\Cref{sec:groups}). %
Additionally, Cauchy theorem states that at last one element would have prime order $p$ if $p$ divides the group order \cite{cauchy1844memoire}, including $p=2$. %
As a consequence, we resort to the compatibility trick (\Cref{sec:id}) to allow the adoption of a group with order that is not exactly a power of $2^8$, but it is a sufficient approximation. %


Let $p$ be a prime number and $\g{p}$ be the set of all $4 \times 4$ unitriangular matrices with entries in the field $\mathbb{F}_p$. %
For the sake of simplicity, we refer to the group $\tup{\g{p},\cdot}$ just as $\g{p}$, where the operator $\cdot$ (or juxtaposition) is the usual matrix multiplication modulo $p$. %
Accordingly, an arbitrary element $a\in\g{p}$ is defined by \Cref{eq:mat}. %
\begin{equation}\label{eq:mat}
a =
\begin{pmatrix}
    1 & e_{12} & e_{13} & e_{14} \\
    0 & 1 & e_{23} & e_{24} \\
    0 & 0 & 1 & e_{34} \\
    0 & 0 & 0 & 1
\end{pmatrix}, \;\;e_{ij} \in \mathbb{F}_p
\end{equation}


Each matrix cell has $p$ possible values, which implies $|\g{p}| = p^6$. %
Therefore, the group UT40.4 has $p=2^{40}\texttt{-}87$ to approximate the 240 bits needed by 40 base-64 digits. %
An advantage of this group is that all non-trivial elements have order $p$ if $p\ge5$. %
The commutativity degree of each group can be calculated from the ratio between its number of conjugacy classes and its order \cite{castelaz2010commutativity,veralopez1995some}, as shown in \Cref{eq:comm}. %
\begin{equation}\label{eq:comm}
\begin{aligned}[b]
   &P_c(\g{p})  &= \frac{2p^3+p^2-2p}{p^6} & \\
   &P_c(\g{2^{32}\texttt{-}5})  &\approx 2.5\cdot10^{-29} &
        \text{\hspace{1cm}\come{UT32.4}} \\
   &P_c(\g{2^{40}\texttt{-}87}) &\approx 1.5\cdot10^{-36} &
        \text{\hspace{1cm}\come{UT40.4}} \\
   &P_c(\g{2^{64}\texttt{-}59}) &\approx 3.2\cdot10^{-58} &
        \text{\hspace{1cm}\come{UT64.4}}
\end{aligned}
\end{equation}
\Cref{eq:map} defines a mapping $m\colon\g{p}\to\mathbb{Z}$ to represent elements according to the intervals introduced in \Cref{sec:id}:
\begin{equation}\label{eq:map}
m(a) = e_{34}*p^5 + e_{12}*p^4 + e_{24}*p^3 + e_{23}*p^2 + e_{13}*p + e_{14}
\end{equation}
where $a$ and $e$ are defined in \Cref{eq:comm}. %
This setting groups commutative elements at the lower range. %
Finally, the digest lexicographic order does not follow the mapping order. %
Each identifier, instead, is composed of one or two different bases to increase compatibility as shown in \ref{appendix}. %
Textual representation (base-16 or base-64) follow the little-endian encoding (most significant digit to the right) to ensure that visually adjacent elements are actually distant from each other according to the mapping cited above. %
This choice favors unique digits at the left of the identifier and also defines reserved elements $\idx{i}$ that are distinct in the part of the identifier corresponding to elements in $\grp{G}\setminus\grp{H}$. %

\subsection{Multi-valued data}\label{sec:multi}
In this section, we present design choices related to the representation of objects and handling of multi-valued data that enable algebraic expressions to cover most data processing workflows. %
Specifically, we target the informational equivalence of possessing a data object and knowing its generating process. %
Thus, identification can be conveniently obtained in two independent ways (aside from history mentioned in \Cref{sec:overview}): %
\begin{itemize}
    \item \textit{from data} - the identifier results from operating the  internal $k$ value-elements of a multi-valued data object $\varphi(u), u\in\mathfrak{V}^k$; %
    \item \textit{from process} - the identifier results from operating the $t$ identifiers of the steps within the composite function $\varphi(s), s\in\mathfrak{F}^t$ that generate $\varphi(u)$. %
\end{itemize}

Except for the reflexive function $\phi$ and asynchronous identifiers (\Cref{sec:async}) which can represent metadata, all functions are expected to affect parts of the data tuple. %
Unneeded values present at the tuple are expected to be ignored by the function and to be kept at the output tuple. %
Maps (lookup tables) are more informative than tuples, and, therefore, they are also considered as a possible type of input data  structure in a separate section (\Cref{sec:map}). %

\subsubsection{Insertion}
A relevant feature of our proposed method is the possibility to simplify the representation of values and functions by considering both as functions. %
A value-element represents a value, but it can also be seen as its insertion function, as illustrated by \textit{Content 3} in \Cref{fig:caixas}. %
The insertion of a value $\varphi(y)$ into a tuple $\tup{\varphi(x)}$ has the same representation regardless of it being interpreted as a function application or as a composite value expression: %
$xy$. %
Therefore, a multi-valued data $\varphi(u), u \in \mathfrak{V}^2$ with two values $\varphi(x)$ and $\varphi(y)$ can be seen as a composition of insertion functions: $u=xy$. %

\subsubsection{Single value creation}\label{sec:conv}
The only difference between insertion and creation, from the output tuple perspective, is that the former extends the tuple to the right, while the latter extends it to the left. %
Such distinct outcomes are due to the fact that the former has no input, while the latter is dependent on the content of the tuple. %
For instance, when the tuple extension results from a function $\varphi(f)$ dependent on input values, a right extension would lead to the problematic implication $xf=xy \implies y=f$. %
Clearly, $y$ would always be the same in this case, even when different input values are provided to $\varphi(f)$. %
On the other hand, the adopted convention correctly leads to a $y$ that depends on $x$:~%
$xf=yx \implies y=xf\inv{x}$. %
For the same reason, (creation) functions must have a non-commuting identifier, i.e., $f\in\grp{G}\setminus\grp{H}$. %

\subsubsection{Multiple values creation}\label{sec:creation}
When a function $\varphi(g)$ generates new values, e.g., $\varphi(z)$ and $\varphi(w)$, it also extends the tuple %
$\varphi(u)=\tup{\varphi(x),\varphi(x)}$ %
to the left as explained in \Cref{sec:conv}:
$\varphi(g)[\varphi(u)] = \varphi(g)[\varphi(xy)] = \tup{\varphi(z),\varphi(w),\varphi(x),\varphi(y)}$. %
At the expression level, the same pattern for single value creation emerges: $ug = xyg = zwxy \implies zw = xyg[xy]^{-1}$. %
However, the elements $z$ and $w$ are not defined. %
Therefore, the element $zw$ must be decomposed in two different factors. %
The general case for multiple values is solved as follows. %

Let $\varphi(v), v\in\mathfrak{V}^{2,3\cdots}$ be a multi-valued data object and $\varphi(f), f\in\mathfrak{F}^1$ be a function that returns $k$ values. %
Their respective elements $x_i \in \grp{G}, 0 \leq i < k-1$ are calculated according to \Cref{eq:multi}, where element $\idx{i} \in \grp{G}$ represents the special $i$th element after the reserved element $\rho$ (\Cref{sec:id}) in digest-based lexicographic order. %
\begin{align}\label{eq:multi}
 \begin{split}
 \left.
    \begin{array}{r@{\mskip\thickmuskip}l}
    \begin{aligned}
        \bm{x} &= vfv^{-1} \\
        \bm{x} &= \prod\limits_{j=1}^{k} x_j \\
        x_{i+1} &= \bm{x}\idx{i}
    \end{aligned}
    \end{array}
    \right\}
    \quad \implies \quad
    \begin{array}{r@{\mskip\thickmuskip}l}
        x_k &= \left(\prod\limits_{j=1}^{k-1} x_j\right)^{-1}\bm{x}
    \end{array}
 \end{split}
\end{align}

\subsubsection{Substitution}
When a function $f$ replaces a value $\varphi(x)$ by $\varphi(x)'$, the new value-element can be derived from the expression, e.g.: %
$xyf = x'y \implies x' = xyfy^{-1}$. %
When a function creates and also replaces multiple values, %
identifiers of replaced and created values are appended to the left of the unchanged elements, creating the expression according to the factorization explained in \Cref{sec:creation}. %
Insertion functions cannot replace existing values. %

\subsubsection{Removal by identity}
From the process point of view, depending on internal data identifiers to be able to perform an operation has no use. %
However, if one has access to the container object, internal values can be removed while keeping consistency between data and process identifiers. %

Value-object removal from the tuple right extremity is trivial. %
It is sufficient to apply the inverse of the chosen element to obtain the expression for the remaining data, e.g.: $xy\inv{y} = x$; or, $\inv{x}xy = y$. %
When the value is in the middle of the tuple, all identifiers from the extremity up to the chosen element must be known so they can be removed and reinserted afterwards. %
For instance, the expression $a=xyzw\inv{[yzw]}zw$ represents the removal of value $\varphi(y)$ from $\varphi(xyzw)$, leading to $a=xzw$. %

Removal by identity is not traceable in history when performed at any place other than the right extremity of the tuple. %
This is due to the nature of process expressions which, aside from the exception of unordered elements, are chronologically ordered from left to right. %
Traceability is only possible while process and history are consistent. %
In this context, one cannot edit the process expression other than right-appending elements (or their inverses). %
Otherwise, history would need to be rewritten to keep consistency, which would clearly undermine the possibility of a reliable audit. %

\subsubsection{Removal by index}\label{sec:deli}
Removal by index does not depend on either content or identity of data objects. %
It is a function object such as any other transformation step. %
The identifier $\sigma_i$ has the parameter $i$ indicating the position in the tuple where a value should be removed. %
Although more realistic than removal by identity, removal by index has a more complex implementation that requires a placeholder $\ph{w}$ for a removed value $\varphi(w), w\in\mathfrak{F}^1$(\Cref{sec:removal}). %
An example is given as follows, showing why this is the case. %

Let the removal of the second element from the tuple $\tup{\varphi(x),\varphi(y)}$ be the desired operation. %
If a placeholder is not used, this would result in the implication $xy\delta_2 = x \implies \delta_2 = \inv{y}$, which is impossible in practice, since only the index is available, not the identifier $y$. %
The function-element $\delta_2$ cannot be defined in terms of identifiers. %
It is an identifier itself, that uniquely identifies a removal operation that could be applied to any tuple with a suitable length. %
On the other hand, when a placeholder is employed, the symbol $\ph{y}$ can be the free variable, as shown by the implication %
$xy\delta_2 = x\ph{y} \implies \ph{y}=y\delta_2$. %

Despite being algebraically possible, removal by identifier and by index are not robust solutions. %
Additionally, operations in tuples in general, including removal by index,  can lead to operational mistakes due to the dependence on how values are ordered along indexes. %
Therefore, when handling tuples, the possibility of removing values could ideally be suppressed depending on the application.  %
A map is a better structure for multi-valued data objects, presented in the \Cref{sec:map}. %

\subsection{Map}\label{sec:map}
Here, a map is a structure where textual keys point to their respective values. %
This enables the use of functions dependent on keys instead of indexes. %
An immediate advantage is to provide protection against arbitrarily ordered objects and to offer more flexibility for the disposition of the entries in a data object. %
Tuple operations described in previous sections still apply for maps. %
Removal by index also applies, provided the map is order-sensitive. %
Removal by name follows the same mechanism as explained in \Cref{sec:deli}. %

Order-insensitive maps have the convenience of having a single identifier for different arrangements of the same content. %
For instance, in most applications, e.g., dealing with original value-elements $x,y,z\in\mathfrak{V}^1$, the inequality $xyz\neq yzx$ brings no advantage. %
This is actually detrimental for original values as it means that each permutation of the entries will have a different, redundant, identifier assigned. %
The only difference between them is the order in which the values were inserted. %
In this context, maps enable the adoption of commutative elements, e.g., $x,y,z\in\grp{H}\setminus\grp{Z}$ to represent a collection of original value-elements $xyz=xzy=zxy=\cdots$ without redundancy while still keeping order in relation to a derived value-element $w\in\grp{G}\setminus\grp{H}$. %
The following equations would hold as expected: %
$xyzw=xzyw=zxyw$; and, $xyzw\neq xwyz\neq wxyz$. %

An important requirement for this solution is to always have the key embedded into original value-elements. %
Otherwise, two different key-value pairs, e.g., $\tup{\hs{x1},\varphi(x)}$ and $\tup{\hs{x2},\varphi(x)}$ would have the same identifier due to the coinciding content $\varphi(x)$. %
Ideally, the embedding should not affect the order-sensitivity of the map entries, but it still have to represent key-value pairs unambiguously. %
Putting the key into the expression as an unordered element would not be a valid solution, e.g.: $[a\gamma(\hs{a})][b\gamma(\hs{b})]$ would be the same as $[b\gamma(\hs{a})][a\gamma(\hs{b})]$. %
One could avoid implying this equality by hashing the key as an ordered element, but it would impose order to the entry, which is also detrimental to the purpose of inserting original values in any order. %
Therefore, an embedding mechanism that preserves the order-sensitivity while keeping key-value relationships is needed. %

We propose a lifting mechanism based on category theory to map both key and value to a non-commutative subgroup $\grp{L}\subset\grp{G}$ where they are combined through an order-sensitive operation and the result is mapped back to $\grp{H}$.
The lifting to the subgroup is formed by switching positions of the matrix cells $e_{23}$ and $e_{12}$ according to the definitions in \Cref{eq:mat}. %
The unlifting is done by switching the same cells. %
For illustration, let $\hat~c$ be the lifted version of $c\in\grp{H}\setminus\grp{Z}$. %
The expression %
$\hat~[\hat~a\hat~\gamma(\hs{a})]\hat~[\hat~b\hat~\gamma(\hs{b})]$ is different from %
$\hat~[\hat~b\hat~\gamma(\hs{a})]\hat~[\hat~a\hat~\gamma(\hs{b})]$ as expected. %
Additionally, this choice avoids keeping several identifiers for the same content in the database, because $x$ refers to the actual content, while $x\gamma(\hs{a})$ refers to the pair key-value. %
The hashing of the key is replaced by a convenient conversion from text to digest detailed in \ref{appendixb}. %

\subsection{Nested data objects}\label{sec:abel}
While flat data structures are sufficient for most applications, nested structures can potentially represent data with unlimited complexity. %
However, a nested data structure such as $\varphi(u)=\tup{\varphi(x), \tup{\varphi(y), \varphi(z)}, \varphi(w)}$ leads to an ambiguous identifier $xyzw$ that could represent the same data in a flat structure: $\varphi(u)\neq\tup{\varphi(x), \varphi(y), \varphi(z), \varphi(w)}$. %
Here, we consider the use of maps as part of the solution. %
The inner map identifier would be generated in a similar manner to that of an original value (\Cref{sec:map}), i.e., embedding the key through lifting: %
$u = x\hat~\set{\hat~[yz]\hat~\gamma(\eta)}w$ where $\eta$ is the key of the inner map for the previous example. %


%

\subsection{Parameterized functions}\label{sec:par}
Functions can be parameterized. %
For instance, a parameter could be a constant indicating the names of input and output values: $xyf_{\tex{x}\to \tex{z}}$. %
Different parameter values usually imply different function identifiers, e.g., $g_\tex{s=1} \neq g_\tex{s=2}$. %
Let the sets $F,G$ be identifiers for all possible parameterizations of the functions $\varphi(f),\varphi(g)$, respectively. %
The set of identifiers for all possible composite functions of $\varphi(g)_*$ with $\varphi(f)_*$ can be expressed by the Cartesian product $H = F \times G$. %
The tuple-expression equivalence (\Cref{sec:notation}) means that $H$ is a set of parameterized $[fg]_*$ expressions that is ultimately a set of identifiers. %
For the sake of convenience, when the different alternatives for a specific position in the set expression includes parameterizations of more than one function, it can be represented by the plus sign, meaning ``union of sets'', e.g.: $(F + G) \times H$. %
Set expressions are useful to describe a sequence of events without knowing the parameter values a priori. %

A current widespread example of   sequence of parameterized functions is a machine learning workflow. %
Each step can be modeled as an identified function: acquisition, cleansing, partition, scaling, enhancement, prediction, evaluation, etc. %
If on one hand, an expression of identifiers can be simplified, resulting in the predicted identifier for the future outcome; on the other hand, a set expression such as the previously mentioned Cartesian product, when simplified, results in a set of outcome identifiers. %

\subsection{Functional Paradigm}\label{sec:fp}
The proposed notation for multi-valued data enables the representation of partial function application \cite{goldberg1987detecting}. %
A function $\varphi(f_{\tex{x},\tex{y},\tex{z}\to\tex{z}})$ expects three values, but can be partially applied to any sequence of values starting from the rightmost value. %
For instance, if $\varphi(f)$ is applied to two values, the identifier $yzf$ represents an intermediate function %
$\varphi(g_{\tex{x}\to\tex{z}})$ %
that could be reused when applied to different $\varphi(x)$ values. %

Moreover, representing a data flow as an expression of identifiers makes it possible
to adopt lazy structures, avoiding duplicate calculations to a great extent. %
When the expression is solved, the resulting identifier is the exact identity of the data object, before any calculation within the flow is started. %
One can also establish the identity of each value in the outcome by following the definitions from the previous sections. %

\subsection{History}\label{sec:hist}
Some applications require all performed steps of a process to be registered in detail for audit purposes. %
In this case, the data structure would keep track of every modification since the creation of the original value-object. %
A list of identifiers is usually sufficient for such a goal. %
This list is updated after each atomic transformation. %
Some items in the list can be sets, when elements from $\grp{H}$ are present, to reflect commuting subexpressions. %
Other items can be tuples, when composite functions are present. %
For instance, the history $\tup{\set{x,y},\tup{g, h},w}$ accounts for: value-elements $x,y$ that are order-insensitive; a function composition $\varphi(h)\circ\varphi(g)$; and, a value insertion $\varphi(w)$.

Dynamic functions, i.e., functions that dynamically apply different substeps according to the input content, can also contribute informatively to the history while keeping a predictable output identifier. %
In such a case, an extra identifier is needed as an adaptor within the expression. %
For instance, $\varphi(f)$ can choose to apply $\varphi(g)$, $\varphi(h)$ or both depending on the input value, e.g., $\varphi(x_1)$ or $\varphi(x_2)$. %
Supposing $\varphi(x_1)$ triggers the application of $\varphi(h)$, and $\varphi(x_2)$ triggers $\varphi(gh)$, adaptors $\bar{f}_1$ and $\bar{f}_2$ would keep consistency between process and history as follows: %
$x_1f = x_1h\bar{f}_1 \implies \bar{f}_1 = h^{-1}f$; and, %
$x_2f = x_2gh\bar{f}_2 \implies \bar{f}_2 = [gh]^{-1}f$.

\subsection{Asynchronous Identifier}\label{sec:async}
Finally, group $\g{2^{40}\texttt{-}87}$ has a maximal Abelian subgroup with order $|\grp{Z}|=2^{40}-87$. %
Identifiers from $\grp{Z}$ always produce the same result regardless of the point at which it is inserted into the expression. %
This property can be useful to represent metadata or any type of step that has no relationship of dependence with any other step. %

\section{Comparison}\label{sec:comparison}
We could not find any proposal in the literature that could be directly compared to our method. %
This is mostly because the originality of the work includes the stated problem/scenario itself, whose applicability is broader than ordinary identification. %
However, we can compare $\g{2^{40}\texttt{-}87}$ to alternative groups. %

Symmetric groups denoted by $\grp{S}_n$ and General Linear groups denoted by $\grp{GL}_{n,q}$ are common examples of finite non-Abelian groups. %
Both have a large proportion of elements with order 2, which makes them unfeasible for our setting. %
Some Dihedral groups denoted by $\grp{D}_{2n}$ have no compatibility gap: %
$\forall \beta \in \mathbb{N}, \exists n \in \mathbb{N}$ such that $\xi_{\beta}(\grp{D}_{2n}) = 0$. %
Special Linear groups denoted by $\grp{SL}_{n,q}$, where elements are matrices whose determinant is equal to the unit of the field $\mathbb{F}_q$, can have a low compatibility gap and low commuting probability as shown in \Cref{tab:comp}. %
\begin{table}
    \caption{Compatibility gap ($\xi_{192}$), commuting probability ($P_c$) and minimum non-trivial element order ($\minord$) for the groups studied as candidates for $\grp{U}$. \textit{Non-integer values are approximated}.}
\label{tab:comp}
    \centering
    \begin{threeparttable}
    \begin{tabular}{lrrr}
    \toprule
        $\grp{G}$ & $\xi_{192}(\grp{G})$ & $P_c(\grp{G})$ & $\minord(\grp{G})$ \\
    \midrule        
        $\grp{S}_{46}$ & 0.123 & $1.92\cdot10^{-53}$ & 2 \\
        $\grp{A}_{46}$ & 0.561 & $1.92\cdot10^{-53}$ & 2 \\
        $\grp{D}_{2^{192}}$ & 0 & $3.98\cdot10^{-59}$ & 2 \\
        $\grp{GL}_{3,2642239}$ & $2.40\cdot10^{-5}$ & $2.94\cdot10^{-39}$ & 2 \\
        $\grp{GL}_{4,4093}$ & 0.012 &  $4.52\cdot10^{-44}$ & 2 \\
        $\grp{SL}_{3,16777213}$ & $1.43\cdot10^{-6}$ &  $2.67\cdot10^{-51}$ & 2 \\
        $\grp{SL}_{4,7129}$ & 0.010 &  $5.80\cdot10^{-47}$ & 2 \\
        $\grp{W}_{2, 7}\times\grp{W}_{2, 13}\times\grp{W}_{2, 23}$ & 0.998 & $2.28\cdot10^{-7\phantom{8}}$ & 7 \\
        $\g{2^{32}\texttt{-}5}$ & $6.98\cdot10^{-9}$ & $2.52\cdot10^{-29}$ & $2^{32}\texttt{-}5$ \\
    \midrule        
        $\g{2^{40}\texttt{-}87}$ & $(4.75\cdot10^{-10})^*$ & $1.50\cdot10^{-36}$ & $2^{40}\texttt{-}87$ \\
    \bottomrule
\end{tabular}
\begin{tablenotes}
\item *Value refers to $\xi_{240}$.
\end{tablenotes}
\end{threeparttable}
\end{table}
However, neither $\grp{D}_{2n}$ nor  $\grp{SL}_{n,q}$ have a high $\minord$. %
Therefore, based on the Lagrange Theorem, we narrowed the options down to non-Abelian $p$-groups. %
As a result, we studied Sylow $p$-subgroups of both $\grp{S}_n$ and $\grp{GL}_{n,p}$. %
If $n=p^k$, then the Sylow $p$-subgroup of $\grp{S}_n$ is the iterated wreath product of the cyclic group $\mathbb{Z}_p$ defined in \Cref{eq:wreath}.
\begin{equation}\label{eq:wreath}
W_{k, p} = \underbrace{\mathbb{Z}_p \wr \mathbb{Z}_p \wr \dots \wr \mathbb{Z}_p}_{k \;\text{times}}
\end{equation}
This group has a minimum non-trivial order equal to $p$. %
However, the group order $|W_{k,p}|=p^{\frac{p^k-1}{p-1}}$ grows exponentially with $p$. %
This implies that low $p$ values are needed, implying in low orders. %
Furthermore, the compatibility gap and the commuting probability are also relatively high as shown in \Cref{tab:comp}. %

Finally, the Sylow $p$-subgroup $\g{p}$ of $\grp{GL}{4,p}$, has a good balance of properties. %
It has a low compatibility gap and commuting probability along with a high minimum non-trivial order. %
Moreover, its implementation is simple as the operations are performed over small matrices. %

\section{Limitations}\label{sec:lim}

Some design choices were made looking for the simplest solution. %
However, some of them were arbitrary and can be changed if any unforeseen limitation is found in the future. %

\subsection{Reference elements}
The choice of $\rho=\hs{\half\halfp}$ and its lexicographic successors $\idx{i}$ as a reference to find factors to compose a multi-valued data creation (\Cref{sec:creation}) is solely based on representational convenience as the symbol \hs{-} is outside the base-62 range of characters and can be regarded as ``internal use only''. %
This choice has high probability of being equivalent to choosing other elements, due to the nature of the group $\g{p}$. %
However, some investigation about the properties of those specific elements could help to identify any side effect when using them, specially regarding corner cases, e.g., when a massive amount of values is created at once. %
Despite being a sequence of successive digests in alphabetical order, the corresponding elements are orders of magnitude distant from each other due to the intentional adoption of the little-endian encoding (\Cref{sec:group}). %

\subsection{Operations}
The \textit{left insertion} convention (\Cref{sec:conv}) implies that a new value-element $z$ produced by the application of a function-element $f$ over $u\in\mathfrak{V}^{2,3\cdots}$ is defined by $z=uf\inv{u}$. %
This leads to the output tuple $\varphi(uf)=\varphi(\tup{uf\inv{u}, u})$. %
At the present, we do not see any ambiguity in $\varphi(uf)$. %
However, we cannot anticipate all outcomes for every possible application. %

The proposed method would be more algebraically flexible if it had two operations ($+$ and $\cdot$) with the distributive property over the finite set of elements. %
Unfortunately, it is theoretically impossible to define a non-commutative finite \textit{division ring} according to the Wedderburn little theorem \cite{bamberg2015completing}. %

\subsection{Group}
We do not expect binary digests are used, but some points are worth of notice if this is the case. %
While the group $\g{2^{40}\texttt{-}87}$ handles 240-bit digests, it can only accept elements with a lexicographic rank lesser than $p^6$, which is not an exact power of 2: %
$2^{239}<p^6<2^{240}$. %
The same constraint applies for creating a commuting element $x\in\grp{H}$. %
Its rank should be lesser than $p^4$. %
Additionally, when importing binary data from legacy systems, the inequality $2^{159}<p^4<2^{160}$ should be taken into account. %
In both cases, the rank should not be less than $p$ (to skip $\grp{Z}$, the center of the group), where elements would commute with any other in $\grp{G}$. %
Analogous inequalities hold for 384-bit digests ($p=2^{64}\texttt{-}59$), whose upper limits are 384 and 256 bits, respectively. %

The first commuting pair of elements is expected, on average, after $2.1\cdot10^{48}$ samples for UT40.4 (\Cref{sec:req}). %
Conversely, expressions are limited in length by \Cref{eq:ambexp}
to avoid ambiguity\footnote{ %
    Avoiding ambiguity only makes sense when a limited set of group elements is effectively used along the existence of an application. Otherwise, every expression is guaranteed to have many algebraically equivalent expressions. %
}. %
For instance, more than $3\cdot10^{15}$ expressions with length $10^7$ will be sampled, on average, before the first ambiguous one is produced according to \Cref{eq:nambexp}. %
This is a comfortable limit as the length for expressions in most applications will probably be in the order of a hundred or less elements. %
Finally, the minimum order $\minord(\g{p})$ limits the number of repeated operations by $p$, which is above $1.1\cdot10^{12}$. %

\subsection{Scenario}
Some limitations are intrinsic of our intended application scenario. %
We adopted the textual digest as the only legitimate format to exchange identifiers between systems. %
This allows a hexdigest such as \hs{ffffffffffffffffffffffffffffffffffffffff} to be used directly as input to our base-64 proposed identification scheme. %
Naturally, the binary representation differs in most cases as the bases are not the same. %
This requires caution if one needs to compare digests in other formats. %
Additionally, this option for textual representation can add a small, mostly negligible, overhead due to conversion between formats. %

Function $\varphi$ is not bijective, i.e., the same object can happen to be identified by more than one identifier. %
While uncommon, two different sequences can produce the same object. %
Since our approach actually identifies the process, not just its current resulting object at hand, it is impossible to guarantee that such a result is unique. %
This is not a problem for any of the intended uses for our scheme, %
but should be taken into account if a specific application produces many similar objects and storing space is a strong priority. %
In such a case, storing the object alongside its hash is recommended to identify it also by content. %
A lookup table linking all its identities to the hash would avoid redundant content entries in the database. %
It is important to notice that the possibility of redundant calculations remains. %
This is the expected behavior and intrinsically unsolvable by any identification schema of choice, given the nature of the scenario. %

\subsection{Other}
The suggested hashing algorithm BLAKE3 can process trillions of petabytes, which is a comfortable limit for current computational needs for a single value. %

\section{Future Work}\label{sec:future}
We intend to explore some new possibilities that $5\times5$ matrices provide, e.g., additional useful subgroups and lower commutability, despite having a considerably lower minimum order as a downside: %
$\minord(\G{p})\approx1.7\cdot10^7$ for 40 digits. %
Regarding the present group ($\g{p}$), which is based on $4\times4$ matrices, we plan to perform an empirical evaluation of the proposed scheme by creating a very small group of identifiers and generating identification expressions exhaustively from it. %
This approach would enable us to discover unexpected artifacts such as a collision/ambiguity rate above acceptable levels under certain circumstances; or, to identify new important limits of the present scheme, among other possibilities. %
Another possible study is to explore formal aspects in detail in a complementary article. %

\section{Conclusion}\label{sec:conc}
In this article, we presented a new scenario for identification systems with an abstract algebra based solution. %
Groups from algebra showed to be specially suitable due to their incidental useful properties for the task of identification. %
They have an identity element that can represent any operation that does not change data such as: storing; logging; or, any monitoring step. %
They are associative, thus representing a composition of several functions as a single identifier. %
Their non-commutative elements can represent chronologically ordered events. %
The commutative subgroups are also useful as they reduce multiplicity of identities for segments within expressions that do not need ordering, such as original values. %
Invertibility can be used to revert an operation or to solve an equation where the source, the process or the outcome are unknown. %
Finally, operations are always closed, i.e., no expression can produce an element outside the finite initial set of identifiers. %

We targeted the scenario where multi-valued data objects are built or transformed by a sequence of functions. %
In this scenario, one can determine the identity of the outcome (or of any value within it) by operating directly on the identifiers from three independent sources of algebraic expressions: %
the process; the history; or, the internal resulting identifiers. %
This flexibility extends current UUID solutions in novel directions that are only marginally explored in this article. %

We presented $\g{p}$ as a concrete group for our scheme. %
Three versions were implemented in software: UT32.4, UT40.4 and UT64.4. %
The proposed method is simple enough to be easily implemented by third parties and also for other programming languages. %

\section{CRediT Author Statement}
\textbf{Davi Pereira-Santos}: Conceptualization, Investigation, Software, Validation, Writing, Review \& Editing. %
\textbf{Gabriel Dalforno Silvestre}: Formal analysis, Software (alternative groups), Writing (part of sections \ref{sec:group} and \ref{sec:comparison}). %
\textbf{André C. P. L. F. Carvalho}: Supervision, Funding acquisition, Resources, Review \& Editing. %

\section{Acknowledgment}
This work was supported by CNPq and FAPESP [grant numbers 2013/07375-0, 2019/01735-0 (CEPID CeMEAI)]. %
We are also grateful for the initial advice from Mark Gritter and Jyrki Lahtonen in some topics of group theory. %

\appendix
\section{Hexdigest Compatibility}\label{appendix}
Two ranges of identifiers are relevant from the compatibility perspective - as shown in \Cref{tab:ranges}. %
The range of hybrid elements is compatible with hexdigests containing one suppressed digit; %
the range of order-sensitive elements is compatible with hexdigests and base-62 digests. %
One digit is suppressed in different positions at the digests of unordered and hybrid elements to avoid ambiguity between elements of the three types, and also to separate digits from different bases. %
The placeholder \_ replaces the suppressed digit and its position (or absence) indicates the type of the element. %
This choice allows to import the same hexdigest as a commutative or non-commutative element. %
Non-commutative elements are prioritized in this design because they have the most general type, i.e., they can be used as commutative ones by mistake without ambiguity while this would not hold for the opposite choice. %
As a minor side effect, commutative elements keep only 31 of their original digits due to suppression. %
\begin{table}
\caption{Identifier ranges according to element type.
All range types contain a subset of hexadecimal digests to provide compatibility.}
\centering
\label{tab:ranges}
\begin{threeparttable}
\begin{tabular}{p{0.16cm}p{0.16cm}p{0.16cm}p{1.2cm}p{0.9cm}lllp{0.1cm}}
\toprule
\multicolumn{3}{p{0.5cm}}{\textbf{Group}}  & \textbf{Element Type} & \textbf{LRI} & \textbf{Digest Range (UT32.4)} & \textbf{Base} \\
\midrule
\multirow{10}{*}{G}&\multirow{7}{*}{H}&\multirow{4}{*}{Z}& identity & $0$ &
\hs{0000000000000000000000000000000000000000} & -- \\
\cline{4-8}  
    &   &   &\multirow{3}{*}{unord.}& \multirow{3}{*}{$[1;p[$} &
\hs{0\_100000000\zfill}$\cdots$ & \multirow{3}{*}{16\_16} \\
    &   &   &           &   &
\hs{0\_fffffffff\zfill}$\cdots$            & &\\
    &   &   &           &   &
\hs{f\_8afffffff\zfill}            & \\
\cline{3-7} 
    &   &   &\multirow{3}{*}{hybrid}& \multirow{3}{*}{$[p;p^{4}[$} &  
\hs{00\_1000000000000000000000000000000000000}$\cdots$   & \multirow{2}{*}{64\_16} \\
    &   &   &           &  &  
\hs{ff\_fffffffffffffffffffffffffffffffffffff}$\cdots$ & \\
    &   &   &     &  &  
\hs{..\_87c2a630003eec7dffff561b0000004aeffff} &  \\
\cline{2-8} 
    &   &   & \multirow{3}{*}{ordered} & \multirow{3}{*}{$[p^4;p^6[$}  &
\hs{1000000000000000000000000000000000000000}$\cdots$ & \multirow{3}{*}{64} \\
    &   &   &  &  &
\hs{ffffffffffffffffffffffffffffffffffffffff}$\cdots$ & \\
    &   &   &    & & \hs{g-8KOjCQREq2Vz8VTc30gLMd..vvX6000ov.....} \\

 \bottomrule
\end{tabular}
\begin{tablenotes}
\item [LRI] Lexicographic Rank.
\item [\grp{Z}] Group center.
\item [\grp{H}] Maximal Abelian subgroup.
\item [\grp{G}] All elements.
\item [--] Not applicable.
\end{tablenotes}
\end{threeparttable}
\end{table}

\section{Key-Digest Conversion}\label{appendixb}
Map keys can be directly used as identifiers within certain constraints. %
Elements in $\grp{H}\setminus\grp{Z}$ have only the two left most digits of the identifier able to accommodate any letter. %
Therefore, keys with more than 2 letters should have the rest of them converted to hexadecimal and be padded with \hs{0}. %
Keys with a single letter are concatenated to \hs{-}. %
\bibliographystyle{plain}
\bibliography{references}
\end{document}